\documentclass[prl,twocolumn,showpacs,preprintnumbers,amsmath,amssymb]{revtex4}

\usepackage{graphicx}
\usepackage{dcolumn}
\usepackage{bm}

\begin{document}

\preprint{APS/123-QED}
\title{Pattern formation with trapped ions}

\author{Tony E. Lee}
\affiliation{Department of Physics, California Institute of Technology, Pasadena, California 91125, USA}
\author{M. C. Cross}
\affiliation{Department of Physics, California Institute of Technology, Pasadena, California 91125, USA}

\date{\today}

\begin{abstract}
Ion traps are a versatile tool to study nonequilibrium statistical physics, due to the tunability of dissipation and nonlinearity. We propose an experiment with a chain of trapped ions, where dissipation is provided by laser heating and cooling, while nonlinearity is provided by trap anharmonicity and beam shaping. The collective dynamics are governed by an equation similar to the complex Ginzburg-Landau equation, except that the reactive nature of the coupling leads to qualitatively different behavior. The system has the unusual feature of being both oscillatory and excitable at the same time. We account for noise from spontaneous emission and find that the patterns are observable for realistic experimental parameters. Our scheme also allows controllable experiments with noise and quenched disorder.
\end{abstract}

\maketitle
Pattern formation is the emergence of structure in a nonlinear medium far from equilibrium \cite{cross93,cross09}. This phenomenon occurs in many settings, including fluids, chemical reactions, plasmas, and biological tissues. In traditional pattern-forming systems, the collective behavior is set by the material properties, and theoretical descriptions are often phenomenological. For example, in the Belousov-Zhabotinsky reaction, the concentrations of chemical reagents oscillate in time and produce traveling waves. It is a complicated reaction involving many intermediate states and rate constants. Hence, it is difficult to experimentally control the behavior of the system.

On the other hand, ion traps allow an unprecedented level of control using optical and electrostatic forces and have led to impressive experiments in quantum computing \cite{leibfried03} and quantum simulation \cite{kim10}. In this paper, we show how ions are also useful for studying pattern formation. The advantage of using ions here is the ability to tune dissipation and nonlinearity \emph{in situ}, thereby having more experimental control and being able to see different effects within the same system.

We show how patterns arise in a chain of ions driven far from equilibrium. The collective dynamics are governed by an equation similar to the complex Ginzburg-Landau equation, which is one of the most studied equations in physics. However, the presence of only reactive coupling in the ion chain leads to novel behavior: while other systems are either oscillatory or excitable, the ion chain can be both at the same time. Our scheme is also useful for studying synchronization and Anderson localization. Our work is motivated by recent experiments on the nonlinear dynamics of single ions: a phonon laser \cite{vahala09,knunz10} and a Duffing oscillator \cite{akerman10}. We note that the model described in this paper is also applicable to an array of nanomechanical resonators \cite{cross04}.

First we describe the proposed experimental setup. A linear Paul trap uses an RF electric field for radial confinement and a DC field for axial confinement \cite{gosh95}. We use a segmented trap, which has many DC electrodes in order to create many trapping regions \cite{kielpinski02}, and thereby make a chain of ions, each in its own potential well. By changing the DC voltages, one can tune the shape of the potential for each ion and thus control the nonlinearity \cite{hensinger06,lin09}. Let $x$ be the axial displacement of an ion from its trap center, $d$ the distance between trap centers, $\omega_o$ the harmonic trap frequency, and $\alpha_o$ the coefficient of the anharmonic quartic term in the trap potential.

\begin{figure}
\centering
\includegraphics[clip,width=3 in, bb=220 450 585 580]{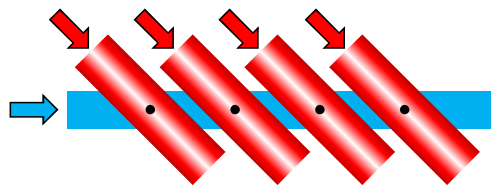}
\caption{\label{fig:setup}Chain of ions, each damped by a red-detuned beam and excited by a blue-detuned beam. The blue beam is along the ion chain, while the red beams are at an angle. The intensity of the red beams is lowest at the trap center. Each ion is in its own anharmonic DC potential well (not shown).}
\end{figure}

We apply near-resonant laser beams to heat and cool each ion (Fig.~\ref{fig:setup}). When the laser frequency is above resonance (blue-detuned), the ion feels an anti-damping force due to the Doppler effect. When the laser frequency is below resonance (red-detuned), the ion feels a damping force \cite{metcalf99}. We apply a blue-detuned beam in the axial direction along the ion chain. For each ion, we also apply a red-detuned beam at an angle $\phi$ with respect to the trap axis. The red beam is shaped so that the intensity is lowest at the trap center. The ion is then heated near the center and increasingly cooled away from the center, so the ion oscillates with a large amplitude, determined by the balance of heating and cooling. The dissipation is easily tuned by changing the beam intensity.

We use a singly-ionized, two-level atom of mass $m$ with a dipole transition of wavelength $\lambda=2\pi/k$ and linewidth $\gamma$. Each red and blue beam has detuning $-\Delta\omega$ and $+\Delta\omega$ and intensity $I_R$ and $I_B$, respectively. Let $I_s$ be the saturation intensity. All beams have the same polarization.


We use low laser intensities so that the ion is unsaturated. We assume the radial motion is cooled near the Doppler limit (due to the projection of the red beams), so that the Doppler shift is due mainly to the axial motion.  For mathematical convenience, we use counter-propagating beams (but in practice one would use single beams). Then the total optical force on an ion is calculated rigorously to be \cite{vahala09},
\begin{eqnarray}
F&=&\frac{8\hbar k^2\gamma^3\Delta\omega}{(\gamma^2+4\Delta\omega^2)^2}\left(-\frac{I_R}{I_s}\cos^2\phi+\frac{I_B}{I_s}\right)\dot{x} \;,
\end{eqnarray}
under the additional assumptions $k|\ddot{x}|\ll\gamma^2/4$ and $k|\dot{x}|\ll\Delta\omega$. To get position-dependent damping, we choose to vary $I_R$ quadratically along the trap axis,
\begin{eqnarray}
I_R(x)&=&\left(\frac{x}{\ell\cos\phi}\right)^2 I_B \;, \label{eq:ir}
\end{eqnarray}
where $\ell$ is the characteristic length of the intensity gradient. The intensity profile does not need to take this form or even be symmetric. In fact, a different profile may lead to interesting higher-order terms in Eq. \eqref{eq:zbartdiscrete} \cite{vansaarloos92}.

The ions are coupled through Coulomb repulsion. If the displacements are small relative to the inter-ion distance $(\ell\ll d)$, the interaction is linear, and the free-space coupling decreases with the cube of the distance. Numerically, we find that interactions farther than the nearest neighbor do not affect the overall dynamics much, so we assume only nearest-neighbor interactions. The equations of motion are,
\begin{eqnarray}\label{eq:eomxchainlinear}
0&=&\frac{d^2}{dt^2}x_n+\omega_o^2 x_n +\alpha_o x_n^3-\mu\left[1-\left(\frac{x_n}{\ell}\right)^2\right]\frac{d}{dt}x_n\nonumber\\
&&+\frac{2k_e e^2}{md^3}\left[(x_n-x_{n-1})+(x_n-x_{n+1})\right] +\chi_n(t),
\end{eqnarray}
$n=1,\ldots,N$, where $k_e$ is the Coulomb constant, $e$ is the proton charge, $\mu$ is the damping coefficient,
\begin{eqnarray}\label{eq:mu}
\mu&=& \frac{8\hbar k^2 \gamma^3\Delta\omega I_B/I_s}{m(\gamma^2+4\Delta\omega^2)^2} \;,
\end{eqnarray}
and $\chi_n(t)$ is the noise. In this scheme, the inherent source of noise is spontaneous emission, since each emission causes a momentum kick $\hbar k$ in a random direction.

We work in the regime where the nonlinearities and interactions are small perturbations to the harmonic motion. We write $x_n(t)=2\ell\,\mbox{Re}[A_n(t) e^{-i\omega t}]$, so that the complex amplitude $A_n$ encodes the slowly varying amplitude and phase of the underlying harmonic oscillations. In the Supplemental Material, we find the amplitude equation,
\begin{eqnarray}
\frac{dA_n}{d\bar{t}}&=&A_n+ib(-2A_n+A_{n-1}+A_{n+1})-(1+ic)|A_n|^2A_n\nonumber\\
&&+[\eta^R_n(\bar{t})+i\sigma^R_n(\bar{t})]A_n+[\eta^B_n(\bar{t})+i\sigma^B_n(\bar{t})]\;, \label{eq:zbartdiscrete}
\end{eqnarray}
\begin{eqnarray}
b=\frac{2k_e e^2}{\nu md^3\omega_o^2},\quad c=\frac{3\alpha_o\ell^2}{\nu\omega_o^2},\quad \nu=\frac{8\hbar k^2\gamma^3\Delta\omega I_B/I_s}{m\omega_o(\gamma^2+4\Delta\omega^2)^2}\;,
\end{eqnarray}
where $\bar{t}=\mu t/2$ is rescaled time, $b$ is the coupling, and $c$ relates how an ion's amplitude affects its frequency. We stress that $b$ and $c$ are directly related to experimental settings. In the absence of coupling, each ion oscillates with amplitude $|A|=1$, which corresponds to an amplitude of $2\ell$ in $x$. The noise functions are due to spontaneous emission and represent scattering by the red beams ($\eta^R$,$\sigma^R$) and blue beams ($\eta^B$,$\sigma^B$),
\begin{eqnarray}
\langle\eta^R_m(\bar{t})\eta^R_n(\bar{t}')\rangle&=&\frac{1}{3}\langle\sigma^R_m(\bar{t})\sigma^R_n(\bar{t}')\rangle=\frac{H\delta(\bar{t}-\bar{t}')\delta_{mn}}{\cos^2\phi}\\
\langle\eta^B_m(\bar{t})\eta^B_n(\bar{t}')\rangle&=&\langle\sigma^B_m(\bar{t})\sigma^B_n(\bar{t}')\rangle=H\delta(\bar{t}-\bar{t}')\delta_{mn}\\
H&=&\frac{\hbar(\gamma^2+4\Delta\omega^2)}{96m\omega_o^2\ell^2\Delta\omega} \;,\label{eq:h}
\end{eqnarray}
where $H$ is a dimensionless measure of the noise.

We now examine the spatiotemporal properties of Eq.~\eqref{eq:zbartdiscrete}, ignoring the effect of noise for now. First note that the equation is symmetric under the transformation $b,c,A_n\rightarrow-b,-c,A_n^*$.
In the continuum limit, we let $A_n\rightarrow A(X)$ and 
\begin{eqnarray}
\frac{dA}{d\bar{t}}&=&A+ib\frac{d^2A}{dX^2}-(1+ic)|A|^2A\label{eq:zbartcontinuum}\;.
\end{eqnarray}
This is similar to the complex Ginzburg-Landau equation (CGLE) \cite{aranson02}, except that the coefficient of $d^2A/dX^2$ is purely imaginary. This is because the Coulomb force is reactive, while the CGLE includes both reactive and dissipative interactions. This greatly modifies the behavior, as seen below.

According to Eq.~\eqref{eq:zbartcontinuum}, a plane wave solution $A(X,\bar{t})=Fe^{i(QX-\omega\bar{t})}$ satisfies $F=1$ and $\omega=bQ^2+c$. By linearizing around this solution \cite{cross09}, we find that the condition for stability is $bc\geq0$, and as long as this is fulfilled, there is no restriction on the wave number $Q$. Since Eq.~\eqref{eq:zbartcontinuum} is in the continuum limit, we expect long wavelength waves (wavelength at least several ions) when $bc\geq0$.

When $bc\leq0$, it turns out that Eq.~\eqref{eq:zbartdiscrete} allows short wavelength waves that are not captured in the continuum limit. Define $\tilde{A}_n=-A_n^*$ for even $n$ and $\tilde{A}_n=A_n^*$ for odd $n$ and consider the continuum limit of the transformed system. A plane wave $\tilde{A}(X)=\tilde{F}e^{i(\tilde{Q}X-\tilde{\omega}\bar{t})}$ satisfies $\tilde{F}=1$ and $\tilde{\omega}=b(\tilde{Q}^2-4)-c$, and the stability condition for long-wavelength waves is $bc\leq0$. A long wavelength wave in $\tilde{A}$ corresponds to a very short wavelength wave in $A$.

Therefore, Eq.~\eqref{eq:zbartdiscrete} has stable plane waves for all values of $b$ and $c$: long wavelength for $bc\geq0$ and short wavelength for $bc\leq0$. This behavior is different from the CGLE, which has stable plane waves for $bc>-1$ and is otherwise chaotic \cite{aranson02}. Another difference is that in the CGLE, only a band of $Q$ is stable.

\begin{figure}
\centering
\includegraphics[width=3.5 in]{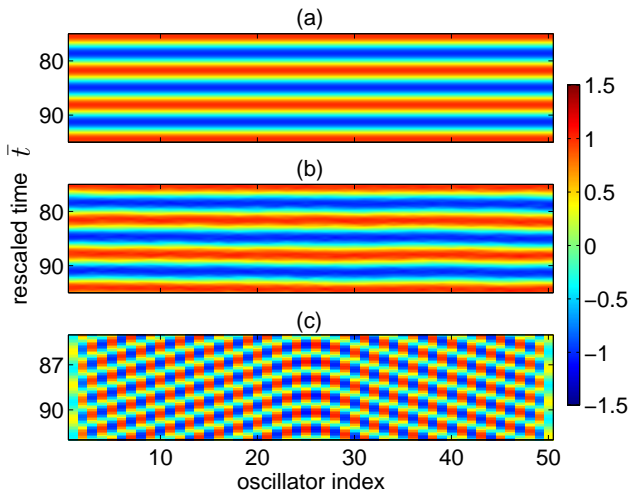}
\caption{\label{fig:n50_wave}Spacetime plot of chain of 50 ions with nearest neighbor interactions and open boundaries. Re $A$ is plotted using the color scale in the side bar. $A$ is the complex amplitude of the underlying harmonic oscillations. (a) $b=1$ and $c=1$ showing uniform phase synchrony. (b) Same, but with the expected noise from spontaneous emission. (c) $b=1$ and $c=-1$ showing anti-phase structure.}
\end{figure}

However, boundary conditions affect the selection of plane waves. With periodic boundary conditions, the wave number ($Q$ or $\tilde{Q}$) is a multiple of $2\pi/N$. Open boundary conditions are simpler to implement and are equivalent to setting $\frac{dA}{dX}=0$ at the boundaries. Thus when $bc>0$, the only allowed plane wave is the $Q=0$ wave, in which the ions are uniformly in-phase (Fig.~\ref{fig:n50_wave}a). (This also occurs in the Belousov-Zhabotinsky reaction in the absence of spatial inhomogeneities \cite{cross09}). One can induce $Q\neq0$ waves by, for example, changing the trap frequency $\omega_o$ of one ion. The $bc<0$ case is different, because open boundary conditions mean setting $\tilde{A}=0$ at the boundaries. The final state is not a pure plane wave but a more complicated structure, in which the ions are almost uniformly anti-phase (Fig.~\ref{fig:n50_wave}c).

\begin{figure}
\centering
\includegraphics[width=3.5 in]{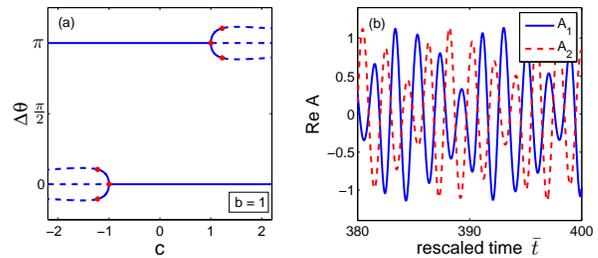}
\caption{\label{fig:n2}System of two oscillators with $b=1$. (a) Bifurcation diagram as $c$ varies. There are supercritical pitchfork bifurcations at $(c,\Delta\theta)=(-1,0)$ and $(1,\pi)$. There are supercritical Hopf bifurcations at $(1.22,2.82)$, $(1.22,3.47)$, $(-1.22,-0.32)$, and $(-1.22,0.32)$, which give rise to stable limit cycles (not shown). Solid and dashed lines denote stable and unstable fixed points, respectively. (b) A limit cycle at $c=1.4$.}
\end{figure}

We now examine the dynamics of two coupled ions, expanding on previous work \cite{kuznetsov09,grau09}. First write Eq.~\eqref{eq:zbartdiscrete} in terms of the real amplitudes $r_1,r_2$ and phase difference $\Delta\theta=\theta_2-\theta_1$, where $A_n=r_n e^{-i\theta_n}$,
\begin{eqnarray}
\frac{d\Delta\theta}{d\bar{t}}&=&(r_2^2-r_1^2)\left(c+\frac{b}{r_1r_2}\cos{\Delta\theta}\right) \label{eq:ddphidbart}\\
\frac{dr_1}{d\bar{t}}&=&(1-r_1^2)r_1+br_2\sin\Delta\theta \label{eq:dA1dbart}\\
\frac{dr_2}{d\bar{t}}&=&(1-r_2^2)r_2-br_1\sin\Delta\theta \label{eq:dA2dbart} \;.
\end{eqnarray}
This system is symmetric under the following transformations: $\{r_1,r_2,\Delta\theta \rightarrow r_2,r_1,-\Delta\theta\}$, $\{c,\Delta\theta \rightarrow -c,\pi-\Delta\theta\}$, and $\{b,\Delta\theta \rightarrow -b,\pi+\Delta\theta\}$.
There are fixed points at $(\Delta\theta,r_1,r_2)=(0,1,1)$ and $(\pi,1,1)$, corresponding to in-phase and anti-phase motion. There is another set of fixed points that correspond to roots of a quartic polynomial of $r_1^2$,
\begin{eqnarray}
0&=&(c^2-1)^2r_1^8-(c^2-1)(c^2-3)r_1^6-(c^2-3)r_1^4 \nonumber\\
 && -(b^2+1)(c^2+1)r_1^2+b^2(b^2+1)\;,
\end{eqnarray}
which may be solved numerically. The bifurcation diagram is quite rich: as $b$ and $c$ change, saddle-node, pitchfork, and Hopf bifurcations appear, disappear, and change criticality. An example is shown in Fig.~\ref{fig:n2}a. For some values of $b$ and $c$, there are supercritical Hopf bifurcations to stable limit cycles, in which the amplitudes and relative phase oscillate (Fig.~\ref{fig:n2}b). The system is at least bistable for $|c|<|b|$, but certain values of $b$ and $c$ have four stable fixed points.

As $N$ increases, there are still in-phase ($\Delta\theta=0$) and anti-phase ($\Delta\theta\approx\pi$) fixed points, although the region of multistability in $bc$ space shrinks. For $|c|\lesssim |b|$, there are also multiple stable limit cycles, in which the entire chain has the same average frequency or the chain is divided into regions of different frequencies.

\begin{figure}
\centering
\includegraphics[width=3.5 in]{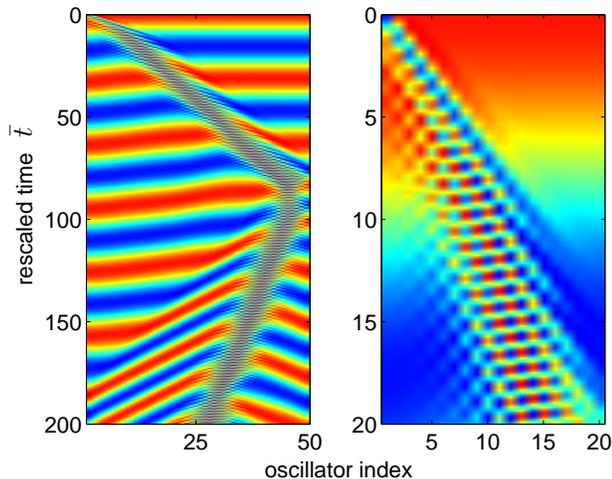}
\caption{\label{fig:n50_pulse}Spacetime plot of Re $A$ for chain of 50 ions showing an excitation pulse for $b=1$ and $c=0.2$. Perturbing the first oscillator beyond a threshold generates a pulse of anti-phase oscillation. The initial conditions were $A_1=-1$ and $A_n=1$ for the rest. The right panel is a zoomed-in view. Color scale is the same as in Fig. \ref{fig:n50_wave}.}
\end{figure}

A large ion chain is excitable in a novel way. An excitable medium has the property that the uniform state is stable to weak perturbations, but a perturbation that exceeds a threshold grows rapidly and then decays. Usually, a medium is either oscillatory or excitable \cite{cross09}. However, the ion chain is both oscillatory and excitable \emph{at the same time}. Suppose that $bc>0$ and the chain is in the $Q=0$ state (in-phase), with $N$ large. We perturb the end ion $A_1$ by $\delta A$. The $Q=0$ state is linearly stable so a small perturbation decays away. But if $\delta A$ is greater than a threshold, it generates a localized pulse of anti-phase oscillations (Fig.~\ref{fig:n50_pulse}). The pulse travels across the system, bouncing off the boundaries until it decays. This type of excitability differs from traditional examples (like neurons and heart tissue), because the pulse is made of an alternating phase structure instead of an increased chemical concentration.

The excitability can be intuitively understood from the fact that when $bc>0$ and $|c|\lesssim |b|$, both in-phase and anti-phase oscillations may be stable for small chains, while only in-phase oscillation is stable for large chains. Thus, a local region within a large chain may be anti-phase for a short amount of time. A mathematical description of this phenomenon is left open for future work.


An important question is whether the patterns described above would be observable for realistic experimental settings that satisfy all the theoretical assumptions. Indeed, we find that the patterns are visible above the noise from spontaneous emission. For example, the ion ${}^{24}\mbox{Mg}^+$ has an $S_{1/2}-P_{3/2}$ dipole transition of wavelength $\lambda=279.6\mbox{ nm}$ and linewidth $\gamma/2\pi=42 \mbox{ MHz}$. Letting $\omega_o/2\pi=100 \mbox{ kHz}$, $I_B/I_s=0.05$, $\Delta\omega=6\gamma$, $\phi=\pi/4$, $\ell=30 \mbox{ $\mu$m}$, $\alpha_o/4\pi^2=10^{15} \mbox{ Hz}^2/\mbox{m}^2$, and $d=500 \mbox{ $\mu$m}$, one finds $b=1.0$, $c=1.1$, and $H=5\times10^{-4}$. Figure \ref{fig:n50_wave}b shows that the $Q=0$ state is clearly visible above the noise. Also, since many experiments are already using large-scale traps for quantum information \cite{kielpinski02}, it should be straightforward to implement our scheme with many ions.

In the experiment, one can measure the amplitude and phase of $A_n$ by recording when and where the ions scatter photons \cite{knunz10,akerman10}. Changes in $A_n$ occur on a time scale $t\sim 1/\mu$, which is much slower than that of the harmonic oscillation ($\sim 1/\omega_o$). Thus, one can observe dynamical effects, such as limit cycles and excitation pulses. The entire $bc$ space can be explored by, for example, tuning the parameters $I_B$ and $\alpha_o$.

It would be interesting to turn up the noise to see what happens. It is known that adding noise to a spatially extended system may have nontrivial effects \cite{sagues07}. However, it is usually experimentally difficult to make the noise fluctuate not only in time but in space as well. In our scheme, the noise for each ion is independent and may be easily tuned by changing the detuning $\Delta\omega$ in Eq.~\eqref{eq:h}. One could observe, for example, noise-induced transitions between stable fixed points in a small chain.

Another interesting use of the tunability is to study the effect of quenched disorder. A previous work studied the mean-field version of Eq.~\eqref{eq:zbartdiscrete} with random harmonic frequencies $\omega_o$ and found that as $b$ and $c$ change, the system undergoes continuous and discontinuous phase transitions between the unsynchronized and synchronized states \cite{cross04,cross06}. It would be interesting to study the lower-dimensional versions. We note that synchronization of disparate oscillators is an important topic throughout science \cite{pikovsky01,acebron05}. Furthermore, when the variance of $\omega_o$ is small, Eq.~\eqref{eq:zbartdiscrete} may be mapped via the Cole-Hopf transformation to the Schr\"{o}dinger equation for a particle in a random potential \cite{sakaguchi88,blasius05}. The resulting pattern formation reflects the phenomenon of Anderson localization.

We thank H. H\"{a}ffner for helpful comments. This work was supported by Boeing and NSF grant DMR-1003337.

\bibliography{pra2010}

\begin{widetext}
\appendix
\section{\label{sec:appendixa}Supplemental Material}

This appendix provides supporting calculations for the main text. In the first section, we derive the amplitude equation using perturbation theory. In the second section, we calculate the noise from spontaneous emission.

\subsection{\label{sec:appendixa1}Derivation of amplitude equation}

Here we use perturbation theory to calculate how the weak nonlinearities and interactions in Eq.~(3) of the main text change the amplitude and phase of the harmonic oscillations on a long time scale. We use the method of averaging with amplitude and phase variables because the noise from spontaneous emission depends on the amplitude. 

First we rescale Eq.~(3) of the main text with $\tau=\omega_o t$ and $y_n=x_n/\ell$,
\begin{eqnarray}
\label{eq:eomytauchain}
0&=&\frac{d^2}{d\tau^2}y_n+y_n+\alpha y_n^3-\nu(1-y_n^2)\frac{d}{d\tau}y_n+D[(y_n-y_{n-1})+(y_n-y_{n+1})]+\zeta_n(\tau) \;,
\end{eqnarray}
where $\alpha=\alpha_o\ell^2/\omega_o^2$, $\nu=\mu/\omega_o$, $D=2k_ee^2/md^3\omega_o^2$, and $\zeta_n(\tau)$ is the noise. Equation \eqref{eq:eomytauchain} describes a chain of van der Pol-Duffing oscillators.

Let $y_n=r_n'(\tau)\cos[\tau+\theta_n(\tau)]$, where $r_n'$ and $\theta_n$ change slowly ($\dot{r}'\ll r'$, $\ddot{r}'\ll\dot{r}'$, $\dot{\theta}\ll 1$, $\ddot{\theta}\ll\dot{\theta}$). Substituting this into Eq.~\eqref{eq:eomytauchain} and keeping the leading terms,
\begin{eqnarray}
0&=&-2\dot{r}_n'\sin(\tau+\theta_n)-2r_n'\dot{\theta}_n\cos(\tau+\theta_n)+\alpha r_n'^3\cos^3(\tau+\theta_n)
+\nu[1-r_n'^2\cos^2(\tau+\theta_n)]r_n'\sin(\tau+\theta_n) \nonumber\\
&&+D[2r_n'\cos(\tau+\theta_n)-r_{n-1}'\cos(\tau+\theta_{n-1})-r_{n+1}'\cos(\tau+\theta_{n+1})]+\zeta_n(\tau) \;.
\end{eqnarray}
Then multiply each equation by $\sin(\tau+\theta_n)$ and integrate over the time interval $[\tau,\tau+\Delta\tau]$, where $\Delta\tau$ is a multiple of $2\pi$,
\begin{eqnarray}
\frac{dr_n'}{d\tau}&=&\frac{\nu}{2}\left(1-\frac{r_n'^2}{4}\right)r_n'+\frac{D}{2}[r_{n-1}'\sin(\theta_{n-1}-\theta_n)+r_{n+1}'\sin(\theta_{n+1}-\theta_n)]+\xi_n^r(\tau) \;. \label{eq:dAdtau}
\end{eqnarray}
Then multiply each equation by $\cos(\tau+\theta_n)$ and integrate similarly,
\begin{eqnarray}
\frac{d\theta_n}{d\tau}&=&\frac{3}{8}\alpha r_n'^2+ \frac{D}{2}\left[2-\frac{r_{n-1}'}{r_n'}\cos(\theta_n-\theta_{n-1})-\frac{r_{n+1}'}{r_n'}\cos(\theta_n-\theta_{n+1})\right]+\xi_n^\theta(\tau) \;. \label{eq:dphidtau}
\end{eqnarray}
These equations describe how $r_n'$ and $\theta_n$ evolve. The noise functions are,
\begin{eqnarray}
\xi_n^r(\tau)&=&\frac{1}{\Delta\tau}\int_\tau^{\tau+\Delta\tau}d\tau'\zeta_n(\tau')\sin(\tau'+\theta_n) \label{eq:xiA}\\
\xi_n^\theta(\tau)&=&\frac{1}{r_n' \Delta\tau}\int_\tau^{\tau+\Delta\tau}d\tau'\zeta_n(\tau')\cos(\tau'+\theta_n) \;. \label{eq:xiphi}
\end{eqnarray}

To put Eqs.~\eqref{eq:dAdtau} and \eqref{eq:dphidtau} in simpler form, rescale time ($\bar{t}=\nu\tau/2$) and amplitude ($r_n=r_n'/2$),
\begin{eqnarray}
\frac{dr_n}{d\bar{t}}&=&(1-r_n^2)r_n+b[r_{n-1}\sin(\theta_{n-1}-\theta_n)+r_{n+1}\sin(\theta_{n+1}-\theta_n)]+\psi_n^r(\bar{t}) \label{eq:dAdtbar}\\
\frac{d\theta_n}{d\bar{t}}&=&c r_n^2 +b\left[2-\frac{r_{n-1}}{r_n}\cos(\theta_n-\theta_{n-1})-\frac{r_{n+1}}{r_n}\cos(\theta_n-\theta_{n+1})\right]+\psi_n^\theta(\bar{t}) \;. \label{eq:dphidtbar}
\end{eqnarray}
where $b=D/\nu$, $c=3\alpha/\nu$, and $\psi_n^r$ and $\psi_n^\theta$ are the rescaled noise functions.
Then write everything in terms of a complex amplitude $A_n=r_ne^{-i\theta_n}$, so that $y_n(\tau)=2\mbox{Re}[A_n(\tau) e^{-i\tau}]$. (We use $e^{-i\theta_n}$ instead of $e^{i\theta_n}$ in order to match up with the sign convention in Ref. \cite{aranson02}.) Then $A_n(\bar{t})$ evolves according to,
\begin{eqnarray}
\frac{dA_n}{d\bar{t}}&=&A_n+ib(-2A_n+A_{n-1}+A_{n+1})-(1+ic)|A_n|^2A_n+\psi_n^A(\bar{t},A_n)\;, \label{eq:dzdbart}
\end{eqnarray}
where $\psi_n^A$ is the complex-valued noise function.

\subsection{\label{sec:appendixa2}Noise from spontaneous emission}

Here we calculate the expected noise from spontaneous emission. When an ion absorbs a photon from a laser, it gets a momentum kick in the direction of the laser, and when it spontaneously emits the photon, it gets a momentum kick in a random direction. Spontaneous emission is the inherent source of noise in our scheme, so we explain how to represent it with the noise term $\psi^A$ in Eq.~\eqref{eq:dzdbart}. 

There are two factors that must be taken into account. First, for the experimental conditions assumed in the text, an ion scatters on the order of one photon per oscillation cycle. Thus, the noise is a sequence of occasional impulses happening at random times. Second, the noise is position dependent due to the intensity gradient of the red beams.

We just consider a single ion, since the noise for each ion is independent and identically distributed. Each scattering event happens at a random time, and the spontaneous emission of a photon causes a momentum kick $\hbar k$ in a random direction. Suppose the ion scatters photons at times $t_n$. Then the noise in Eq.~(3) of the main text is
\begin{eqnarray}
\chi(t)&=&\frac{\hbar k}{m} \sum_n \delta(t-t_n)q_n \;,
\end{eqnarray}
where $q_n$ is a random variable (with variance $\sigma_q$) for the projection of a momentum kick along the trap axis. Each kick is independent ($\langle q_jq_k\rangle=\delta_{jk}$). For simplicity, we assume that the emission is isotropic ($\sigma_q^2=1/3$), although there is a slight anisotropy relative to the laser direction \cite{stenholm86}.  

With the assumptions on experimental parameters given in the text, the scattering rate $\Gamma$ may be calculated rigorously from the Optical Bloch Equations \cite{cohen92,vahala09},
\begin{eqnarray}
\Gamma(x)&=&\frac{\gamma^3}{I_s}\left[
\frac{I_R(x)}{\gamma^2+4\Delta\omega_R^2}
+\frac{I_B}{\gamma^2+4\Delta\omega_B^2} \right] \;.
\end{eqnarray}
The first and second terms correspond to scattering by the red and blue beams, respectively. Note that $\Gamma$ depends on position and is independent of velocity to first order.

After rescaling ($\tau=\omega_o t$ and $y=x/\ell$) to get Eq.~\eqref{eq:eomytauchain}, the noise is
\begin{eqnarray}
\zeta(\tau)&=&\frac{\hbar k}{m\omega_o\ell} \sum_n \delta(\tau-\tau_n)q_n \;, \label{eq:zeta}
\end{eqnarray}
and the scattering rate becomes,
\begin{eqnarray}
\tilde{\Gamma}(y)&=&\tilde{\Gamma}_R(y)+\tilde{\Gamma}_B \label{eq:ry} \\
\tilde{\Gamma}_R(y)&=&\frac{1}{\omega_o}\left(\frac{y}{\cos\phi}\right)^2\frac{I_B}{I_s}\frac{\gamma^3}{\gamma^2+4\Delta\omega^2} \\
\tilde{\Gamma}_B&=&\frac{1}{\omega_o}\frac{I_B}{I_s}\frac{\gamma^3}{\gamma^2+4\Delta\omega^2} \;,
\end{eqnarray}
where we have used the intensity relation given in Eq.~(2) of the main text.

To calculate the amplitude noise $\xi^r$, plug Eq.~\eqref{eq:zeta} into Eq.~\eqref{eq:xiA},
\begin{eqnarray}
\xi^r(\tau)&=&\frac{\hbar k}{\Delta\tau m\omega_o\ell} \sum_{\tau<\tau_n<\tau+\Delta\tau} q_n\sin(\tau_n+\theta) \;, \label{eq:xiAsum}
\end{eqnarray}
where $\tau_n$ is the time of a scattering event. Since the damping is weak, the ion scatters on the order of one photon in an oscillation cycle ($\Gamma\sim\omega_o/2\pi$), so there is significant time between scattering events. This means that the phase of oscillation at which a scattering event occurs is approximately uncorrelated with the phase of the next event. Each scattering event has a random projection and phase. Thus, the sum in Eq.~\eqref{eq:xiAsum} is over independent samples of the random variable $w_n=q_nu_n$, where $u_n=\sin\tau_n$ (ignoring the unimportant phase offset $\theta$ for now).

Now we find $w_n$'s distribution $\rho_w$. The intensity gradient of the red beams causes them to scatter more at certain phases within a cycle, while the blue beams scatter uniformly. Thus, $\rho_w$ is actually a weighted average of red and blue components. First we find the distribution of scattering times $\tau_n \mbox{ (mod }2\pi)$ from the intensity profiles,
\begin{eqnarray}
\rho_\tau(\tau)&=&
\left\{\begin{array}{l l}
\frac{1}{\pi}\cos^2\tau & \quad\mbox{red}\\
\\
\frac{1}{2\pi} & \quad\mbox{blue}
\end{array}\right.  \;,\label{eq:rhotau}
\end{eqnarray}
since $y=r'\cos\tau$. Thus the distribution of $u_n=\sin\tau_n$ is
\begin{eqnarray}
\rho_u(u)&=&\rho_\tau\left|\frac{d\tau}{du}\right|\\
&=&
\left\{\begin{array}{l l}
\frac{2}{\pi}\sqrt{1-u^2} & \quad\mbox{red}\\
\\
\frac{1}{\pi}\frac{1}{\sqrt{1-u^2}}  & \quad\mbox{blue}
\end{array}\right.  \;,
\end{eqnarray}
for $|u|\leq1$. Since we assume isotropic spontaneous emission, the distribution of the projection $q_n$ is $\rho_q(q)=1/2$ for $|q|\leq1$.
Then the distribution of $w_n=q_nu_n$ is
\begin{eqnarray}
\rho_w(w)&=&\int_{-1}^1 du \int_{-1}^1 dq \;\rho_u(u)\rho_q(q)\delta(w-uq)\\
&=&
\left\{\begin{array}{l l}
\frac{2}{\pi}\left[-\sqrt{1-w^2}+\log\frac{1+\sqrt{1-w^2}}{|w|}\right] & \quad\mbox{red}\\
\\
\frac{1}{\pi}\log\frac{1+\sqrt{1-w^2}}{|w|}  & \quad\mbox{blue}
\end{array}\right.  \;,
\end{eqnarray}
for $|w|\leq1$. The variance of $w_n$ is
\begin{eqnarray}
\sigma_w^2&=&
\left\{\begin{array}{l l}
\frac{1}{12} & \quad\mbox{red}\\
\\
\frac{1}{6}  & \quad\mbox{blue}
\end{array}\right. \;.
\end{eqnarray}

To find the phase noise $\xi^{\theta}$, plug Eq.~\eqref{eq:zeta} into Eq.~\eqref{eq:xiphi},
\begin{eqnarray}
\xi^\theta(\tau)&=&\frac{\hbar k}{r'\Delta\tau m\omega_o\ell} \sum_{\tau<\tau_n<\tau+\Delta\tau} q_n\cos(\tau_n+\theta) \;, \label{eq:xiphisum}
\end{eqnarray}
and go through the same process to find the variance of $v_n=q_n\cos\tau_n$,
\begin{eqnarray}
\sigma_v^2&=&
\left\{\begin{array}{l l}
\frac{1}{4} & \quad\mbox{red}\\
\\
\frac{1}{6}  & \quad\mbox{blue}
\end{array}\right. \;. \label{eq:rhov}
\end{eqnarray}
Although $w_n$ and $v_n$ come from the same scattering event, they are statistically uncorrelated because $\langle\sin\tau_n\cos\tau_n\rangle=0$.

We let the time interval $\Delta\tau$ be large enough to include many scattering events but smaller than the characteristic time scales in Eqs.~\eqref{eq:dAdtau} and \eqref{eq:dphidtau}. We average the scattering rate $\tilde{\Gamma}(y)$ in Eq.~\eqref{eq:ry} over $\Delta\tau$ to find the time-averaged scattering rates of the red beams ($\bar{\Gamma}_R$) and blue beams ($\bar{\Gamma}_B$),
\begin{eqnarray}
\bar{\Gamma}_R(r')&=&\frac{1}{2\omega_o}\left(\frac{r'}{\cos\phi}\right)^2\frac{I_B}{I_s}\frac{\gamma^3}{\gamma^2+4\Delta\omega^2}\\
\bar{\Gamma}_B&=&\frac{1}{\omega_o}\frac{I_B}{I_s}\frac{\gamma^3}{\gamma^2+4\Delta\omega^2} \;.
\end{eqnarray}
$\bar{\Gamma}_R$ depends on $r'$ due to the intensity gradient of the red beam.
Then the amplitude and phase noises are Gaussian and described by,
\begin{eqnarray}
\langle\xi^r(\tau)\xi^r(\tau')\rangle&=&\left(\frac{\hbar k}{m\omega_o\ell}\right)^2\left(\frac{1}{12}\bar{\Gamma}_R+\frac{1}{6}\bar{\Gamma}_B\right)\delta(\tau-\tau') \label{eq:xiAxiA}\\
\langle\xi^\theta(\tau)\xi^\theta(\tau')\rangle&=&\frac{1}{r'^2}\left(\frac{\hbar k}{m\omega_o\ell}\right)^2\left(\frac{1}{4}\bar{\Gamma}_R+\frac{1}{6}\bar{\Gamma}_B\right)\delta(\tau-\tau') \;.\label{eq:xiphixiphi}
\end{eqnarray}
They are uncorrelated with each other: $\langle\xi^r(\tau)\xi^\theta(\tau')\rangle=0$.

After rescaling ($\bar{t}=\nu\tau/2$, $r=r'/2$) to get Eqs.~\eqref{eq:dAdtbar} and \eqref{eq:dphidtbar}, the noises become,
\begin{eqnarray}
\langle\psi^r(\bar{t})\psi^r(\bar{t}')\rangle&=&\frac{1}{2\nu}\left(\frac{\hbar k}{m\omega_o\ell}\right)^2\left(\frac{1}{12}\bar{\Gamma}_R+\frac{1}{6}\bar{\Gamma}_B\right)\delta(\bar{t}-\bar{t}') \label{eq:psiApsiA}\\
\langle\psi^\theta(\bar{t})\psi^\theta(\bar{t}')\rangle&=&\frac{1}{2\nu r^2}\left(\frac{\hbar k}{m\omega_o\ell}\right)^2\left(\frac{1}{4}\bar{\Gamma}_R+\frac{1}{6}\bar{\Gamma}_B\right)\delta(\bar{t}-\bar{t}') \;. \label{eq:psiphipsiphi}
\end{eqnarray}
Again, $\langle\psi^r(\bar{t})\psi^\theta(\bar{t}')\rangle=0$. Finally, the complex-valued noise in Eq.~\eqref{eq:dzdbart} is,
\begin{eqnarray}
\psi^A(\bar{t},A)&=&[\eta^R(\bar{t})+i\sigma^R(\bar{t})]A+[\eta^B(\bar{t})+i\sigma^B(\bar{t})]\\
\langle\eta^R(\bar{t})\eta^R(\bar{t}')\rangle&=&\frac{H}{\cos^2\phi}\delta(\bar{t}-\bar{t}')\\
\langle\sigma^R(\bar{t})\sigma^R(\bar{t}')\rangle&=&\frac{3H}{\cos^2\phi}\delta(\bar{t}-\bar{t}')\\
\langle\eta^B(\bar{t})\eta^B(\bar{t}')\rangle&=&\langle\sigma^B(\bar{t})\sigma^B(\bar{t}')\rangle=H\delta(\bar{t}-\bar{t}')\\
H&=&\frac{\hbar(\gamma^2+4\Delta\omega^2)}{96m\omega_o^2\ell^2\Delta\omega} \;,
\end{eqnarray}
where $H$ is a measure of the noise, and we have simplified using Eq.~(4) of the main text. The noise functions for the red beams ($\eta^R$,$\sigma^R$) and blue beams ($\eta^B$,$\sigma^B$) are all uncorrelated with each other. The noise from the red beams increases with amplitude and causes more phase noise than amplitude noise.

\end{widetext}

\end{document}